\documentstyle[aps,epsf,prl,multicol]{revtex}

\begin{document}
\draft

\title{\hfill{\small SOGANG-ND 78/99} \\
\vspace{0.5cm}
Phase Synchronization with Type-II Intermittency in 
       Chaotic Oscillators}

\author{Inbo Kim$^{\dag,\ddag}$, Chil-Min Kim$^{\dag,\ast}$, Won-Ho
  Kye$^{\dag}$, and Young-Jai Park$^{\dag,\ddag}$}
\address{
\hspace{0.1in}$^{\dag}$National Creative Research Initiative Center for
    Controlling Optical Chaos,\\
\hspace{0.2in}Department of Physics, Pai Chai University, Seogu,
Taejon, 302-735, Korea \\
$^{\ddag}$Department of Physics, Sogang University, Seoul,  121-742, Korea}
\maketitle

\begin{abstract}
We study the phase synchronization (PS) with type-II intermittency showing
$\pm 2 \pi$ irregular phase jumping behavior before the PS transition
occurs in a system of two coupled hyperchaotic R\"{o}ssler oscillators. The
behavior is understood as a stochastic hopping of an overdamped
particle in a potential which has $2 \pi$-periodic minima. We
characterize it as type-II intermittency with external noise through
the return map analysis. In $\epsilon_{t} < \epsilon < \epsilon_{c}$
(where $\epsilon_{t}$ is the bifurcation point of type-II
intermittency and $\epsilon_{c}$ is the PS transition point in
coupling strength parameter space), the average length of the time
interval between two successive jumps follows $\langle l\rangle \sim
\exp(\mid\!\epsilon_{t} - \epsilon\!\mid^{2})$, which agrees well with
the scaling law obtained from the Fokker-Planck equation.           

\end{abstract}

\begin{multicols}{2}

Synchronization is one of the basic phenomena ubiquitously found in
physical, chemical, biological and physiological systems. In the
classical sense, synchronization of {\em periodic} self-sustained
oscillators is usually defined as locking of the phases, 
$n\theta_{1} - m\theta_{2} = const$, due to weak interaction 
while the amplitudes can be quite different. This phenomenon has been
quite well studied and has witnessed a lot of practical applications
in various engineering fields~\cite{bl}. 

Recently, the notion of synchronization has been extended to 
coupled {\em chaotic} oscillators (i.e., individual
oscillators are chaotic without coupling). One of the remarkable
developments is the observation of phase synchronization (PS)
phenomenon in a system of two mutually coupled nonidentical
self-sustained chaotic oscillators~\cite{ps,pik}. The phenomenon is
analogous to synchronization of periodic oscillators where only the
phase locking is a matter of importance. Above a critical strength of
coupling, suitably defined phases of two chaotic oscillators lock each
other and synchronize, while their amplitudes remain chaotic and
uncorrelated with each other. Also it was found that the phase
difference $\phi$ between two oscillators increases with an
intermittent sequence of $2 \pi$ jumps before the PS transition
occurs. It means that phase slips occur from time to time and the
phase difference changes by $2 \pi$ during a rather small interval of time. 

The intermittent behavior and its scaling properties near the PS
transition in a coupled R\"{o}ssler system was studied by several
authors~\cite{lee,kye}. They provided an
explanation for the phase jumps by reducing the original system into a
simplified model describing an overdamped particle sliding in a
``noisy wash-board potential''. Also by studying the scaling rules of
the jumping behavior, they found that the phenomenon is related
with type-I intermittency~\cite{man} in the presence of
noise~\cite{lee,kye}. So far, this has been known to be the only route
to the PS transition in two coupled self-sustained chaotic oscillators,
while nothing forbids other types of intermittency to exist.    

In this paper, we report another route to PS transition exhibiting
$\pm ~ 2 \pi$ jumps, which is characterized by the type-II
intermittency with external noise. We consider the following two
coupled hyperchaotic R\"{o}ssler oscillators (HRO's),
 
\begin{eqnarray}
\stackrel{.}{x}_{1,2}&=&-\Omega_{1,2}{y}_{1,2} - z_{1,2} +
\epsilon ({x}_{2,1} - {x}_{1,2}), \nonumber \\
\stackrel{.}{y}_{1,2}&=&\Omega_{1,2} {x}_{1,2} + 0.25 {y}_{1,2} +
{w}_{1,2}, \nonumber \\ 
\stackrel{.}{z}_{1,2}&=&3.0 + {x}_{1,2} {z}_{1,2},\\
\stackrel{.}{w}_{1,2}&=&-0.5 {z}_{1,2} + 0.05 {w}_{1,2} + \epsilon
({w}_{2,1} - {w}_{1,2}), \nonumber 
\end{eqnarray}
where two variables $x$ and $w$ are coupled and the subscripts 1 and 2
refer to each of the oscillators.
Here $\Omega_{1,2} = 1.0  \pm  \Delta\Omega/2$\ \ is the overall
frequency of each chaotic oscillator, and $\epsilon$ is the
coupling strength. 

To observe PS we must define a suitable phase related to this system.
Since the phase portrait of a hyperchaotic R\"{o}ssler oscillator in the
$x$-$y$ plane explicitly shows a rotational trajectory around a
center, $(x_{0}, y_{0})$ as shown in Fig.~1, the phase can be defined
in relation with this rotation, i.e., $x_{1,2}$ and $y_{1,2}$ are
transformed into polar variables $A_{1,2}$ and $\theta_{1,2}$ around
the center of the rotation where $A_{1,2} = \sqrt{(x_{1,2} - x_{0})^{2} +
(y_{1,2} - y_{0})^{2}} $ and $\theta_{1,2} =
\tan^{-1}\left(\frac{y_{1,2} - y_{0}}{x_{1,2} - x_{0}}\right)$. 
With these variables, Eq.~(1) can be rewritten as follows:
\begin{eqnarray}
\stackrel{.}{A}_{1,2}&=& 0.25A_{1,2}\sin^2\theta_{1,2} -
z_{1,2}\cos\theta_{1,2} + \tilde{w}_{1,2}\sin\theta_{1,2} \nonumber \\
&+& \epsilon(A_{2,1}\cos\theta_{2,1}\cos\theta_{1,2} -
A_{1,2}\cos^2\theta_{2,1}), \nonumber \\
\stackrel{.}{\theta}_{1,2}&=&\Omega_{1,2} +
0.25\sin\theta_{1,2}\cos\theta_{1,2} \nonumber \\
&+& \frac{z_{1,2}}{A_{1,2}}\sin\theta_{1,2} +
\frac{\tilde{w}_{1,2}}{A_{1,2}}\cos\theta_{1,2} \nonumber \\
&-& \epsilon\left(\frac{A_{2,1}}{A_{1,2}}\cos\theta_{2,1} -
\cos\theta_{1,2}\right)\sin\theta_{1,2},   \\ 
\stackrel{.}{z}_{1,2}&=&3.0 + ({A}_{1,2}\cos\theta_{1,2} + x_{0}){z}_{1,2},
\nonumber \\
\stackrel{.}{w}_{1,2}&=&-0.5{z}_{1,2} + 0.05 {w}_{1,2} + \epsilon
({w}_{2,1} - {w}_{1,2}). \nonumber 
\end{eqnarray}
where $\tilde{w}_{1,2}$ is a shorthand for $w_{1,2} +
x_{0}\Omega_{1,2}$ and $x_{0} = -16.0$ and $y_{0} = 0.0$ are used.
By solving Eq.~(1) or (2) numerically, the phase
difference ($\phi = \theta_{1} - \theta_{2}$) of the two oscillators is
obtained as shown in Fig.~2. Note that the onset of PS occurs near
$\epsilon_{c} = 0.18$~\cite{comment}. Before the onset of PS, the
smaller the coupling is the more frequently $2 \pi$ jumps occur like
the case of the R\"{o}ssler system. But in our system, the phase
difference jumps by $2 \pi$ upwards or downwards irregularly as shown
in Fig.~2, while in the R\"{o}ssler case it jumps only upwards
monotonously. In this respect, our system should be distinguished from the
R\"{o}ssler system qualitatively~\cite{ps,lee,kye}. 

Here, we are interested in this new type of
phase jumping behavior and would like to uncover the basic
structure behind it.     
For this purpose, the following equation for the phase
difference $\phi$ is obtained from Eq.~(2):
\begin{eqnarray}
\frac{d\phi}{dt} 
&=& \Delta\Omega + (0.25 + \epsilon)(\sin\theta_{1}\cos\theta_{1} -
\sin\theta_{2}\cos\theta_{2}) \nonumber \\
&+& \frac{z_{1}}{A_{1}}\sin\theta_{1} -
\frac{z_{2}}{A_{2}}\sin\theta_{2} + \frac{\tilde{w}_{1}}{A_{1}}\cos\theta_{1} -
\frac{\tilde{w}_{2}}{A_{2}}\cos\theta_{2} \nonumber \\
&-& \epsilon\left( \frac{A_{2}}{A_{1}}\sin\theta_{1}\cos\theta_{2} -
  \frac{A_{1}}{A_{2}}\cos\theta_{1}\sin\theta_{2} \right).  
\end{eqnarray}
The r.h.s. of Eq.~(3) can be rewritten in terms of $\phi$ by grouping
relevant terms, and then we obtain the following phase equation
similar to that describing phase locking of periodic
oscillators in the presence of noise~\cite{ps,piko,stra}: 
  
\begin{eqnarray}
\frac{d\phi}{dt} &=& \Delta\Omega + \alpha \sin\phi + \beta
\sin\frac{\phi}{2} + \xi,  
\end{eqnarray}
where  
\begin{eqnarray*}
\alpha &=& ( 0.25 + \epsilon)\cos(\theta_{1} + \theta_{2})  
- \frac{\epsilon}{2} \left( \frac{A_{2}}{A_{1}} + \frac{A_{1}}{A_{2}}
\right), \\
\beta &=& 2\left\{ \frac{z_{1}}{A_{1}}
  \cos \left( \frac{\theta_{1} + \theta_{2}}{2} \right) 
- \frac{\tilde{w}_{1}}{A_{1}} \sin \left( \frac{\theta_{1} + \theta_{2}}{2}
\right)\right\},  \\
\xi &=& \left( \frac{z_{1}}{A_{1}} - \frac{z_{2}}{A_{2}} \right)
\sin\theta_{2} + \left( \frac{\tilde{w}_{1}}{A_{1}} -
  \frac{\tilde{w}_{2}}{A_{2}} \right) \cos\theta_{2}\\
&-& \frac{\epsilon}{2} \left( \frac{A_{2}}{A_{1}} -
  \frac{A_{1}}{A_{2}} \right) \sin(\theta_{1} + \theta_{2}).  
\end{eqnarray*}
Note that there are two time scales, i.e., the fast one, $2
\pi/\Omega_{1,2}$, related with the frequency of each individual
oscillator in Eq.~(2), and the slow one, $2\pi / \Delta\Omega$,
originating from the frequency mismatch in Eq.~(3) which is the
characteristic time of $\phi$ dynamics. Since $\alpha$ and $\beta$ and
$\xi$ are fast fluctuating pieces compared with the time scale of
$\phi$ variable, qualitative features of $\phi$ dynamics can be
revealed after averaging $\alpha$ and $\beta$ over the slow
time scale, while $\xi$ is left intact as an external noise in Eq.~(4):
      
\begin{equation}
\frac{d\tilde{\phi}}{dt} = F(\tilde{\phi}, \epsilon) + \xi, 
\end{equation}
where 
\begin{equation}
F(\tilde{\phi}, \epsilon) = \Delta\Omega + \langle\alpha\rangle
\sin\tilde{\phi} + \langle\beta\rangle \sin\frac{\tilde{\phi}}{2}.  
\end{equation}  
In Eq. (5) and (6), by $\tilde{\phi}$ we mean a new phase variable
which simulates original $\phi$ dynamics.     
Here $\langle\alpha\rangle$, $\langle\beta\rangle$ represent
averaged values for a long enough time interval which is order of $2
\pi/\Delta\Omega$. This equation describes an overdamped particle
moving in a potential under the influence of external noise $\xi$,
where the potential $V(\tilde{\phi},\epsilon)$ is defined by
$F(\tilde{\phi}) = - \frac{dV(\tilde{\phi})}{d\tilde{\phi}}$~\cite{lee,stro}.
This averaged potential is obtained by integrating the force $F(\tilde{\phi},
\epsilon)$ with respect to $\tilde{\phi}$, i.e.,
$V(\tilde{\phi},\epsilon) = - \int^{\tilde{\phi}} 
F(\tilde{\phi}',\epsilon)d\tilde{\phi}'$ up to an arbitrary
integration constant:    

\begin{equation}
V(\tilde{\phi}, \epsilon) = -\Delta\Omega\tilde{\phi} + \langle\alpha\rangle
\cos\tilde{\phi} + 2 \langle\beta\rangle \cos\frac{\tilde{\phi}}{2}.  
\end{equation} 

Figure 3 (a), (b), and (c) show time series of $\phi$ at $\epsilon = 0.16$,
the corresponding force $F$, and the potential $V$ respectively. 
As shown in Fig.~3 (c), the existence of $2 \pi$-periodic minima is the main
feature of the potential related with the phase jumping dynamics. Note
that the overall slope of this potential is negligible due to the small
constant force term  $\Delta\Omega$, so that the ``wash-board
potential'' picture~\cite{lee} is not applicable to studying the
bifurcation related with the intermittency in this system. With this
potential structure, however, we can qualitatively explain the phase
jumping dynamics as a hopping of an overdamped particle between $2
\pi$-periodic minima in the potential through the {\em stochastic
process} driven by the external noisy force $\xi$.
In Fig.~3 (c), when a particle is in a potential well, it can jump
randomly to the nearest neighboring potential wells since the barrier
heights of both sides of the well are similar. When the well is not
deep, the particle resides there for a short time interval, but it
resides there longer when the well is deep. This behavior can be
easily observed in the time series of $\phi$.  

To determine the type of intermittency explicitly, we obtain the
return map of $\phi$ from the original Eq.~(2) by finding
$\phi_{n}$ at every $2\pi$ rotation of phase $\theta_{1}$. Figure 4
(a) shows a schematic relation among force, potential, and the
corresponding return map. In Fig.~4 (b), the return map shows many
cells separated by $2 \pi$ corresponding to $2 \pi$-periodic minima in
the potential structure. The trajectory in the return map moves from
one cell to the neighboring cells irregurally in accordance with $\pm
 2 \pi$ jumps. To see the structure more clearly, we obtain the
return map of $\tilde{\phi}$ from the averaged Eq.~(5) in one cell at
every $2 \pi$ rotation of phase $\theta_{1}$ when $\xi = 0$. The solid
curves in Fig.~4 (b), (c), and (d) are the results and they quite well 
follow the one obtained directly from the original system (2). This means
that the averaged Eq.~(5) properly describes the main features of $\pm
 2 \pi$ phase jumping dynamics. From the shape of the curves, it is
clear that the return map is actually that of type-II
intermittency~\cite{man,kim}. In Fig.~4 (c) the center is a stable
fixed point which becomes unstable when the slope at this point equals
$1.0$ at the bifurcation point $\epsilon_{t} = 0.038$ as shown in
Fig.~4 (d). The trajectory can escape from the center to the nearest
neighboring stable fixed points located each at $\pm ~ 2 \pi$ distance
away through the stochastic process. Therefore the scaling of the
average laminar length of $\phi$\ (here the laminar length is the time
elapsed between two successive $\pm 2 \pi$ jumps) is expected to
follow that of type-II intermittency in the presence of external noise.    

In order to verify the above conjecture, we compare the average
laminar length scaling relation obtained numerically from Eq.~(2) with the
analytical result obtained from the Fokker-Planck equation (FPE).  
The local Poincar\'{e} map of type-II intermittency with
external noise is described by the following difference
equation~\cite{stro,kim,hir,ott}:
\begin{equation}
x_{n+1} = (1 + \epsilon) x_{n} + a x_{n}^3 + \sqrt{2 D}\xi_{n},
\end{equation}
where $a$ is a positive arbitrary constant, and $D$ is the dispersion
of Gaussian noise $\xi_{n}$.
It is well known that the above map (8) can be approximated to
the backward FPE in the long laminar region~\cite{risk}:
\begin{equation}
\frac{\partial G(x,t)}{\partial t} = - V'(x)\frac{\partial
  G(x,t)}{\partial x} + D\frac{\partial^2 G(x,t)}{\partial^2 x}
\end{equation}
where $G(x,t)$ is the probability density of particle at $(x,t)$. 
The scaling relation for
the average escaping time can be derived analytically from this
FPE. According to the analytical estimation made in our
recent work~\cite{bok}, the average laminar length scales as follows
when $\epsilon_{t} < \epsilon$:
\begin{equation}
\langle l\rangle \sim \langle l\rangle_{0} \exp(\mid\!\epsilon_{t} -
\epsilon\!\mid^{2}).  
\end{equation}
       
Now, we perform numerical simulation to obtain the scaling law of the
average laminar length from Eq.~(2) and the results are shown
in Fig.~5. First we get the value of
$y$-intercept, $ \ln \langle l\rangle_{0}$, in Fig.~5 (a) to determine
the scaling exponent by the linear regression.  The slope of the
regression line in Fig.~5 (b) is very close to 2.0 within $\pm ~ 5 ~
\%$ error, which agrees with the scaling exponent in
Eq.~(10). Therefore our conjecture based on the return map analysis is
verified and strengthened by this remarkable agreement with the result
obtained from the FPE. Also this agreement in turn justifies our averaging out
the fast fluctuating factors $\alpha$ and $\beta$ in Eq.~(5), which
marginally contribute as multiplicative perturbations in phase jumping
dynamics.


In conclusion, we have firstly found the type-II intermittency route
to the PS transition in a system of two coupled self-sustained
HRO's. In this system, the phase
difference between two oscillators exhibits apparently irregular
$\pm ~ 2 \pi$ jumps near the onset of the PS transition.
Furthermore, we have identified this novel behavior
as type-II intermittency with external noise through the return map
analysis and the FPE approach.
 
We thank Dr. S. Rim and D. U. Hwang for valuable discussions. 
This work was supported by the Project of Creative Research
Initiatives, Korea Ministry of Science and Technology.
Y.-J. Park 
acknowledges the support from the Korean
Council for University Education for 1999 Domestic Faculty Exchange.

\begin{figure}[htbp]
\epsfxsize=8cm
\epsfbox{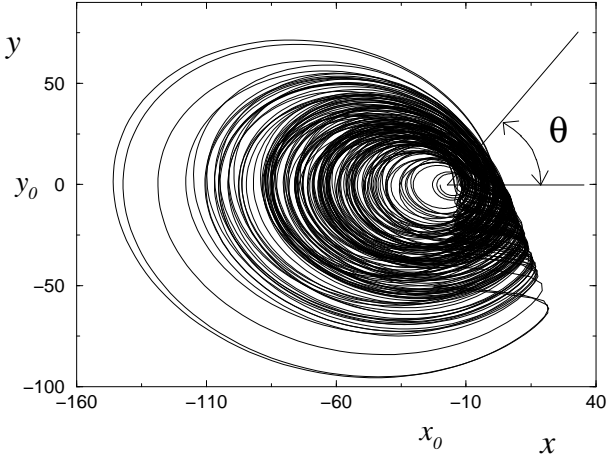}
\narrowtext
\vspace{0.5cm}
\caption{The phase portrait of one of two coupled HRO's projected on
  the $x$-$y$ plane.
  Initial values for HRO's 1 and 2 are $x_{1}(0) =
  -20.0$, $ y_{1}(0) = z_{1}(0) = 0.0$, $w_{1}(0) = 15.0$ and
  $x_{2}(0) = -20.1$, $ y_{2}(0) = z_{2}(0) = 0.0$, $w_{2}(0) = 15.1$
  respectively. Eq.~(1) is numerically solved by using a fourth-order
  Runge-Kutta method with the parameter mismatch $\Delta\Omega =
  0.001$, and the same values are used throughout this Letter.} 
\end{figure}

\begin{figure}[htbp]
\epsfxsize=8cm
\epsfbox{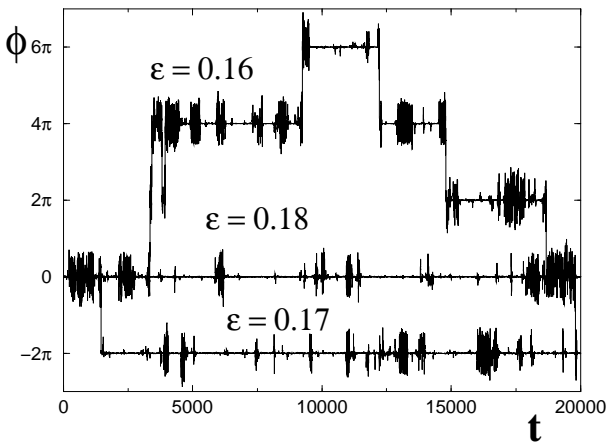}
\narrowtext
\vspace{0.5cm}
\caption{Time series of phase difference $\phi$ in 
  two coupled HRO's for various values of (a)
  $\epsilon = 0.16$, (b) $\epsilon = 0.17$, and (c) $\epsilon =
  0.18$. $\pm ~ 2 \pi$ phase jumps are shown for the cases of (a) and (b),
  while for (c) the PS transition seems to be achieved for the observed time
  scale.} 
\end{figure}

\begin{figure}[htbp]
\epsfxsize=8cm
\epsfbox{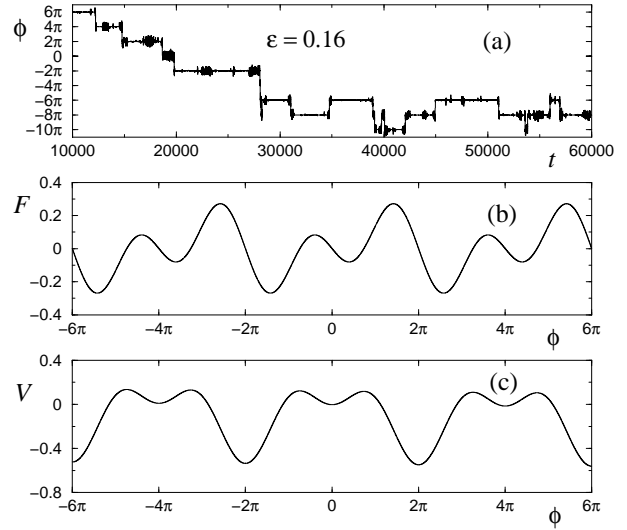}
\narrowtext
\vspace{0.5cm}
\caption{(a) Time series of phase jumping dynamics, (b) the
  time-averaged force $F(\phi)$, and (c) the potential $V(\phi)$ at
  $\epsilon = 0.16$.} 
\end{figure}

\begin{figure}[htbp]
\epsfxsize=8cm
\epsfbox{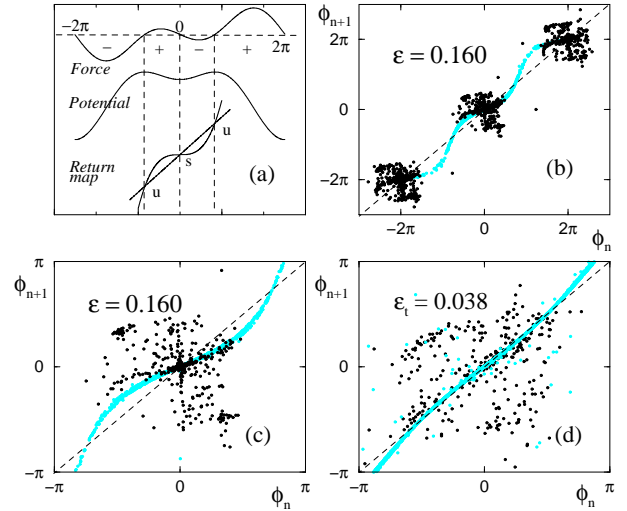}
\narrowtext
\vspace{0.5cm}
\caption{(a) schematic relation among force, potential, and return
  map; (b) return map (black dots) of original system (1)  and return
  map (gray dots) from the averaged Eq.~(4); (c) and (d) magnified return
  maps around one potential well with $\epsilon = 0.16$ and
  $\epsilon_{t} = 0.038$ respectively. Note that the tangent line at
  the center stable fixed point becomes tangential to the diagonal
  line in (d).} 
\end{figure}        

\begin{figure}[htbp]
\epsfxsize=8cm
\epsfbox{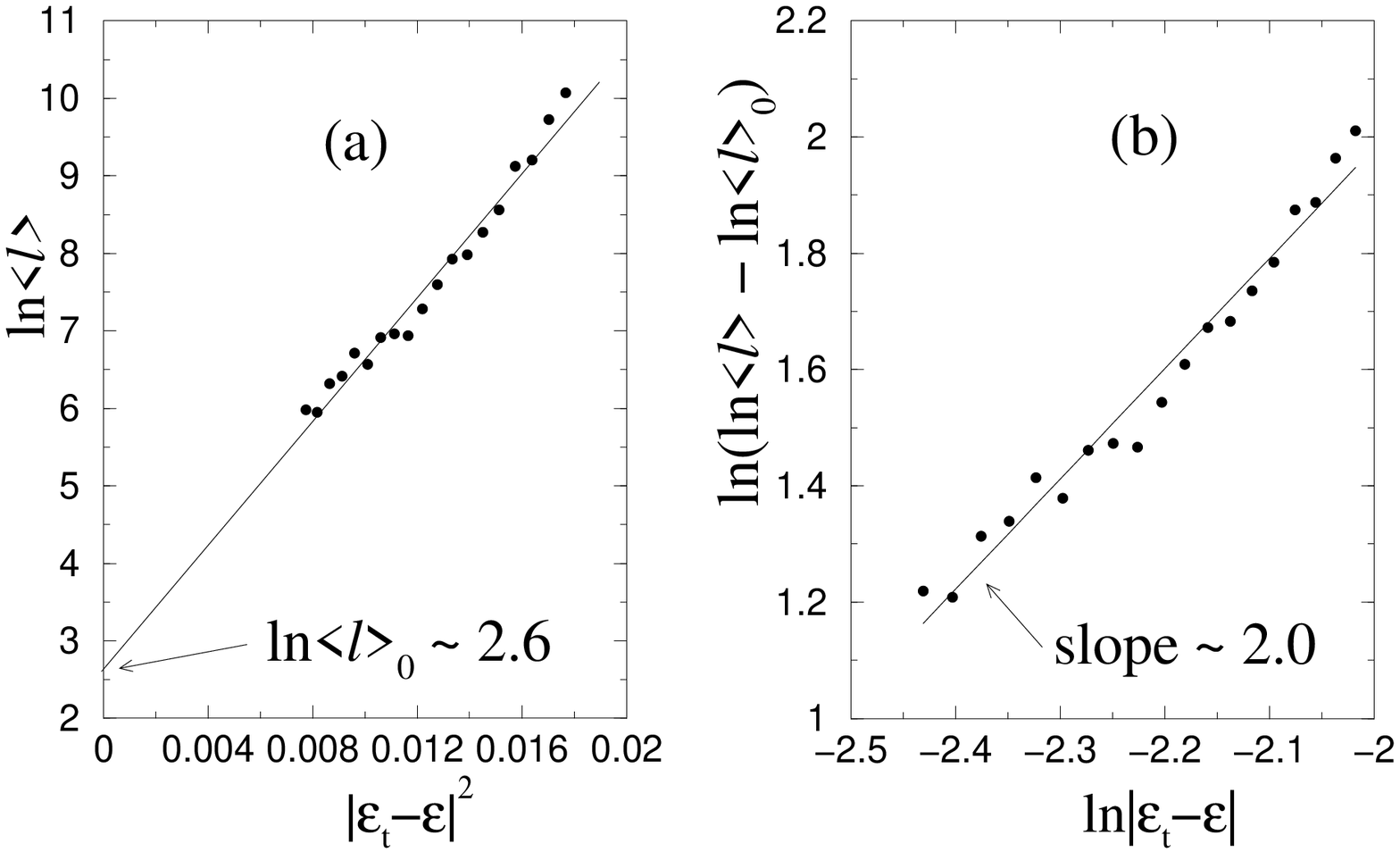}
\narrowtext
\vspace{0.5cm}
\caption{Scaling relation: average laminar length $\ln\langle
  l\rangle$ vs $\mid\!\epsilon_{t} - \epsilon\!\mid^{2}$ in (a) and
  $\ln(\ln\langle l\rangle - 2.6)$ vs. $\ln\!\mid\!\epsilon_{t} -
  \epsilon\!\mid$ in (b). Note that the slope in (b) is approximately
  2.0 within $\pm ~ 5 ~ \%$ error in agreement with the analytic
  estimation from the FPE.} 
\end{figure}        

\end{multicols}

\begin{thebibliography}{99}
\bibitem{comment} {We cannot unambiguously determine the border of
    this transition. The transition to the synchronous state is always
    smeared and observed only as a tendency, or as a temporary event on
    some finite time intervals. According to Eq.~(10), there
    doesn't exist the onset point of PS because the average laminar
    length increases exponentially. But here the point $\epsilon_{c} =
    0.18$ is practically what can be observed in numerical simulation
    within the limit of computation time.}
\end{thebibliography}
\end{document}